\documentclass[a4paper]{PoS}
\usepackage{booktabs,graphicx,siunitx,xspace}
\usepackage{subfigure}
\usepackage{wrapfig}
\usepackage{epsfig}
\usepackage{epstopdf}
\newcommand{\Msol}{M\ensuremath{_\odot}\xspace}
\newcommand*\aap{A\&A}

\newcommand*\apj{ApJ}
\newcommand*\apjl{ApJ}

\newcommand*\araa{ARA\&A}

\newcommand*\mnras{MNRAS}

\newcommand*\nar{New A Rev.}
\newcommand*\nat{Nature}

\let\oldbibliography\thebibliography
\renewcommand{\thebibliography}[1]{%
  \oldbibliography{#1}%
  \setlength{\itemsep}{0pt}%
}

\title{Gamma-ray lines from SN2014J}

\ShortTitle{Gamma-ray lines from SN2014J}

\author{\speaker{Thomas Siegert} and Roland Diehl\\
        Max Planck Institut f\"ur extraterrestrische Physik, D-85748 Garching, Germany\\
        E-mail: \email{tsiegert@mpe.mpg.de},
\email{rod@mpe.mpg.de}}

\abstract{On 21 January 2014, SN2014J was discovered in M82 and found to be the closest type Ia supernova (SN Ia) in the last four decades. INTEGRAL observed SN2014J from the end of January until late June for a total exposure time of about 7~Ms. SNe Ia light curves are understood to be powered by the radioactive decay of iron peak elements of which $^{56}$Ni is dominantly synthesized during the thermonuclear disruption of a CO white dwarf (WD). The measurement of $\gamma$-ray lines from the decay chain $^{56}$Ni$\rightarrow$$^{56}$Co$\rightarrow$$^{56}$Fe provides unique information about the explosion in supernovae. Canonical models assume $^{56}$Ni buried deeply in the supernova cloud, absorbing most of the early $\gamma$-rays, and only the consecutive decay of $^{56}$Co should become directly observable through the overlaying material several weeks after the explosion when the supernova envelope dilutes as it expands.\\
Surprisingly, with the spectrometer on INTEGRAL, SPI, we detected $^{56}$Ni $\gamma$-ray lines at 158 and 812~keV at early times with flux levels corresponding to roughly 10\% of the total expected amount of $^{56}$Ni, and at relatively small velocities. This implies some mechanism to create a major amout of $^{56}$Ni at the outskirts, and at the same time to break the spherical symmetry of the supernova. One plausible explanation would be a belt accreted from a He companion star, exploding, and triggering the explosion of the white dwarf.\\
The full set of observations of SN2014J show $^{56}$Co $\gamma$-ray lines at 847 and 1238~keV, and we determine for the first time a SN Ia $\gamma$-ray light curve. The irregular appearance of these $\gamma$-ray lines allows deeper insights about the explosion morphology from its temporal evolution and provides additional evidence for an asymmetric explosion, from our high-resolution spectroscopy and comparisons with recent models.}

\FullConference{10th INTEGRAL Workshop: "A Synergistic View of the High Energy Sky" - Integral2014,\\
		15-19 September 2014\\
		Annapolis, MD, USA }

\begin{document}

\section{Introduction}
Supernovae of type Ia are thought to be standard candles for the cosmic distance ladder at distances of more than $\approx$100~Mpc. This is based on the uniformity of most measured light curves in the optical bands~\cite{Goobar2011_SNIa}. It is not clear what the progenitor system of a SN Ia is, except that a CO white dwarf and its thermonuclear disruption must be part of it. There is more than one possible branch for a WD to enter this final fate. One possibility is a giant star companion delivering the additional fuel for the WD to ignite in its center when reaching its Chandrasekhar limiting mass~\cite{Hillebrandt2000_SNIa,Li2011_SNIa}. It can also undergo a triggered explosion by an accreted He layer which is detonating on the surface~\cite{Hillebrandt2013_SNIa}. These scenarios are called single degenerate (SD) and have been the consensus for several years. Recently, additional observations and increasing model sophistication revealed more diversity and offered plausible alternatives. A broader range of binary systems and more methods of igniting a white dwarf, independent of its mass, manage to destabilize the degenerate matter, e.g. by accretion flow instabilities~\cite{Guillochon2010_SNIa} or by merging two white dwarves (double degenerate, DD)~\cite{Hillebrandt2013_SNIa,Rosswog2009_SNIa,Fink2010_SNIa,Pakmor2012_SNIa}. The common product of all the different models is the radioactive isotope $^{56}$Ni which is produced in nuclear fusion reactions at high densities, once the WD is ignited~\cite{Nomoto1997_SNIa}. This explosive nuclear burning competes with the expansion of the star, resulting in some outer parts of the WD being burnt towards intermediate mass elements instead of to iron group nuclei, or even left unburnt as carbon and oxygen~\cite{Mazzali2007_SNIa}. Depending on the models and by indirect estimations~\cite{Arnett1982_SNIa}, the typical amout of $^{56}$Ni produced is expected to be between 0.1 and 0.9~\Msol embedded in about 0.5 to 0.9~\Msol of other material~\cite{Mazzali2007_SNIa,Stritzinger2006_SNIa}.\\
Up to several weeks, the SN cloud should be opaque even to penetrating $\gamma$-rays from the $^{56}$Ni decay with a characeteristic life-time of 8.8~days~\cite{Isern2008_SNIa}. During this time, the outer gas absorbs the radioactive energy input and re-radiates it at lower energies from UV through IR, with such absorption becoming less efficient as the SN expands. Neither the explosion dynamics nor the evolution towards explosions from WD properties and from their interactions with their companion stars can easily be assessed by observations in these wavelengths~\cite{Roepke2012_SNIa}. Only pre-explosion data brought some insights towards the nature of the binary companion star~\cite{Li2011_SNIa,Bours2013_SNIa,Ruiz2004_Tycho,Kerzendorf2009_Tycho,Schaefer2012_SNR0509}, and from its interactions with the SN~\cite{Bloom2012_SN2011fe,Shappee2013_SN2011fe,Goobar2014_SN2014J}.\\
Deeper insights can be obtained by direct measurements of $\gamma$-rays from the $^{56}$Ni decay chain $^{56}$Ni$\rightarrow$$^{56}$Co$\rightarrow$$^{56}$Fe, leaking out of the expanding SN after several weeks. According to most recent models, the consecutive decay from $^{56}$Co to $^{56}$Fe with a characteristic time scale of 111~days should show a maximum in $\gamma$-ray emission brightness at days 60-100 after the explosion, declining thereafter due to the radioactive decay of $^{56}$Co~\cite{Nomoto1984_SNIa,Khokhlov2001_SNIa,Woosley1994_SNIa,Livne1995_SNIa,Hillebrandt2000_SNIa,Mazzali2007_SNIa,Isern2011_SNIa,Milne2004_SNIa,Hillebrandt2013_SNIa,Dessart2014_SNIa,The2014_SN2014J}. The brightness of $^{56}$Co decay $\gamma$-rays during the time after the explosion is determined by the initial amount of available $^{56}$Ni and how it is distributed within the expanding SN. This is the phase in which the rise and fall of the $\gamma$-ray line intensity provides unique information on the type of the explosion and the structure of the SN and thus gives hints on the progenitor system.\\
SN2014J is a unique opportunity to investigate these aspects using the spectrometer SPI~\cite{Vedrenne2003_SPI} on INTEGRAL~\cite{Winkler2003_INTEGRAL}
because it is the closest SN Ia in the last 40 years~\cite{Cao2014_SN2024J}. It was accidentally discovered by an undergraduate class lead by Steve Fossey on January 22, 2014~\cite{Fossey2014_SN2014J}, and recognised as a type Ia explosion from early spectra~\cite{Cao2014_SN2024J}. It occured in the nearby starburst galaxy M82 at a distance of $\approx$3.3~Mpc~\cite{Foley2014_SN2014J}, and is seen only through major interstellar material in the host galaxy along the line-of-sight, causing rather large reddening~\cite{Goobar2014_SN2014J}. This naturally involves difficulties in measuring the proprietary SN brightness and low-energy spectra at high precision.\\
INTEGRAL started to observe SN2014J on January 31~\cite{Kuulkers2014_SN2014J}, and pointed to it until June 26
accumulating a total exposure of $\approx$7~Ms. This corresponds to observations of SN2014J during days 16.3 to 164.0 after the inferred explosion date of January 14, UT $14.75~\pm~0.21$~\cite{Zheng2014_SN2014J}.\\
Using our new analysis method for determination of the dominating instrumental background~\cite{Diehl2014_SN2014J_Ni,Diehl2014_SN2014J_Co}, for the first time we find SN Ia $\gamma$-ray lines at 158 and 812~keV originating from the decay of $^{56}$Ni. At only 20~days after the explosion, the detection itself and the spectral shapes are unexpected and uncover asymmetries in the explosion. INTEGRAL's follow-up observations have been used to provide a SN Ia $^{56}$Co $\gamma$-ray light curve. This, at first glance, conforms to most models. But with SPI's unique spectral resolution ($\approx$2~keV), the observed $^{56}$Co $\gamma$-ray line shapes provide evidence for an asymmetry in SN2014J and also supports a double detonation scenario.

\section{A new analysis method}
Generally, in our spectroscopic analysis we fit scaling factors of proposed sky model intensity distributions plus models of instrumental background (BG) components to the set of spectra accumulated during the observations. This resembles linear regression of specific, given time sequences, incorporating the instrumental (imaging) response function. Hence, data $d_k$ of energy bin $k$ are modeled as a linear combination of the $N_I$ sky model components $M_{ij}$, to which the instrument response matrix $R_{jk}$ is applied, and the $N_B$ BG components $B_{jk}$:
\begin{equation}
d_k = \sum_j R_{jk} \sum_{i = 1}^{N_\mathrm{I}} \theta_i M_{ij} +
\sum_{i = N_\mathrm{I} + 1}^{N_\mathrm{I} + N_\mathrm{B}} \theta_i
B_{jk}\label{eq:model-fit}
\end{equation}
i.e. the comparison is performed in data space with respect to time\footnote{The term "time" is somewhat misleading since one and the same pointing can be observed at different times. However, for simplicity, "time" is used in order to describe any means of "a particular pointing at time t".}, which consists of the counts per energy bin $k$ measured in each of SPI's detectors $j$ for each single telescope pointing as part of the complete observation.\\
For several years, the common approach was to treat the count rate in each energy bin separately, thus not further using spectral response information of the instrument. The BG was determined empirically by using other instrumental rates as a kind of radiation monitor for the BG stimulating events. Since these rates do not exactly resemble the BG count rate over a longer period of time, additional scaling parameters had to be introduced. Moreover, the BG shows strong variations with energy, and for this reason these methods need to be calibrated for each energy bin separately.\\
Our new approach for determining the BG count rate is based on detailed spectral fits in the energy region of interest. We thus investigate the nuclear reaction physics happening inside the satellite
by studying the raw spectra for each detector and each orbit that is part of the analysed data set.
A three-day integration is used to avoid statistical limitations in the BG count prediction per pointing and helps to identify even minor spectral features. We confirmed independently that spectral BG features do generally not change over such periods.
\begin{figure}
\vspace{-20pt}
  \centering
  \includegraphics[width=0.8\linewidth,clip=true,trim=0.25cm 0.25cm 0.25cm 0.25cm]{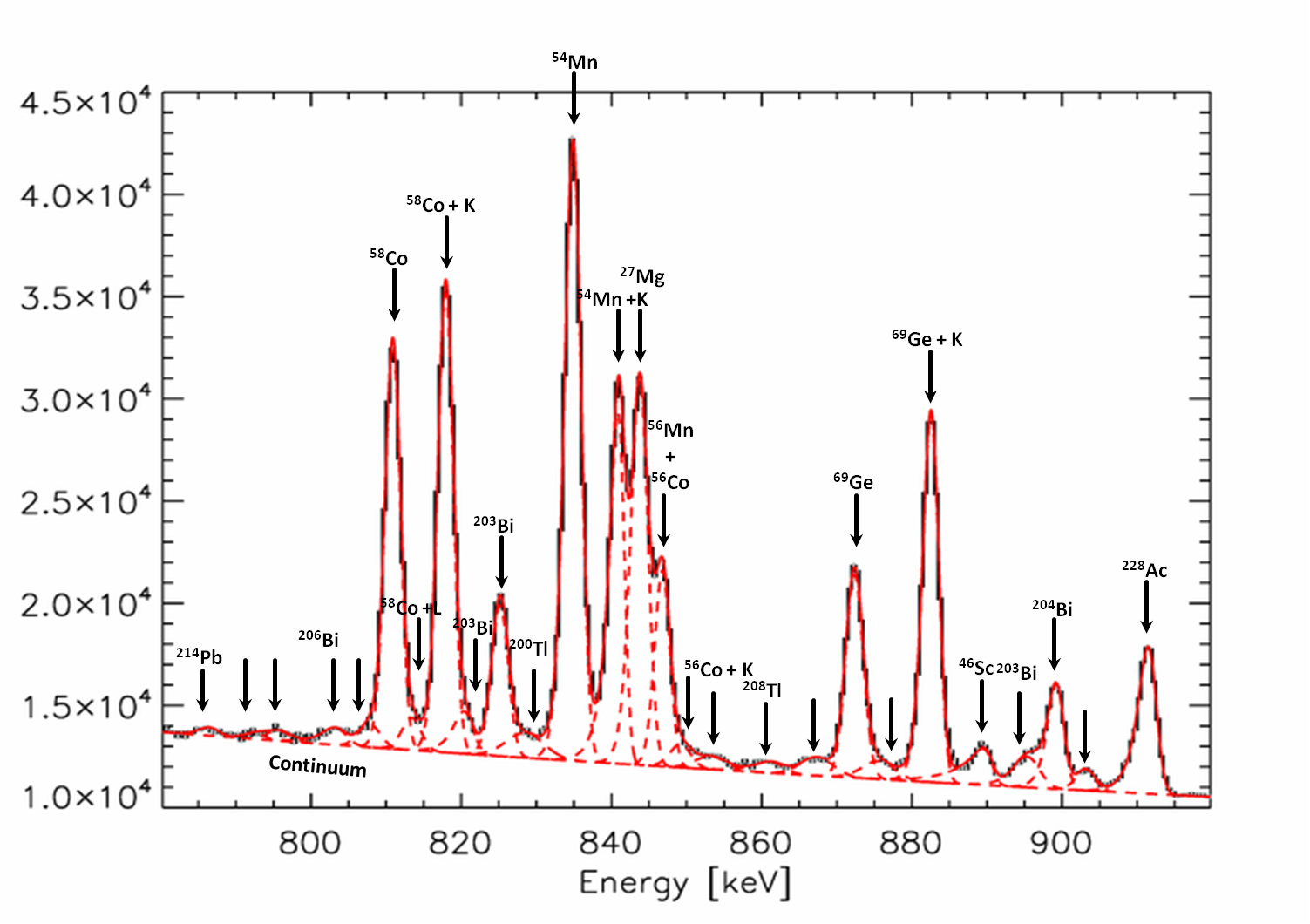}
  \caption{Raw orbit-integrated count spectrum of the SPI camera. Shown are the photon counts in the energy band 780 - 920~keV per half keV energy bin in black and superimposed a spectral fit in red, identifying 28 different components: A $\gamma$-ray continuum, described by a power-law, and 27 instrumental $\gamma$-ray lines (20 already mentioned in e.g.~\cite{Weidenspointner2003_SPI}, indicated with arrows).
   }
  \vspace{-10pt}
  \label{fig:fit_with_iso}
\end{figure}
We determine spectral shape paramters of each instrumental feature (see figure~\ref{fig:fit_with_iso}) for all detectors by fitting the spectra using a Metropolis-Hastings algorithm in a Monte-Carlo Markov-Chain and fix these spectral templates during one orbit. The spectral response function used for fitting the data is represented by a symmetric Gaussian function, convolved with an exponential tail function towards lower energies which incorporates the degradation due to cosmic bombardment in each detector. It has four fit parameters per detector and orbit in each $\gamma$-ray line~\cite{Kretschmer2011_PhD}. Since each detector shows a slightly different degradation at a particular point in time, and in addition reacts differently to a certain BG source inside the satellite (being more or less illuminated than another detector), a unique and independent time-sequence of BG counts can be predicted. Long-term investigations showed that the relative intensity of a particular spectral BG feature (e.g. a $\gamma$-ray line originating from the anti-coincidence shield) in one detector with respect to the mean intensity of the same feature in all detectors (detector ratios) is constant in time, and thus maximally independent of possible sky signals. This spectral-shape and detector-ratio constancy is imprinted onto the short-term
variations while the mean amplitude (over all detectors) is still allowed to vary.\\
Together with plausible candidate celestial source models, in the case of SN2014J one single point source, the scaling parameters $\theta_i$ in equation~(\ref{eq:model-fit}) for both, BG and sky, are then determined by a maximum likehood fit simultaneously, setting up the BG and celestial spectra.

\section{$^{56}$Ni $\gamma$-ray measurements}
\begin{figure}[h!tbp]
\vspace{-20pt}
\subfigure{\includegraphics[width=0.49\textwidth]{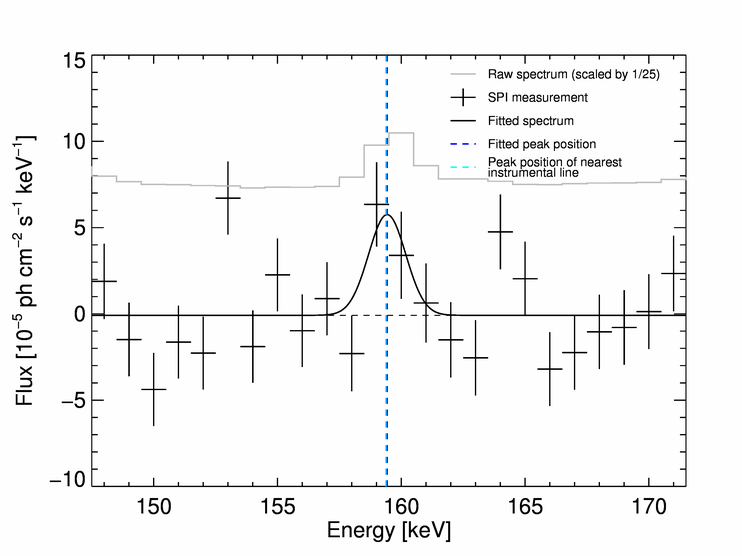}}\hfill
\subfigure{\includegraphics[width=0.49\textwidth]{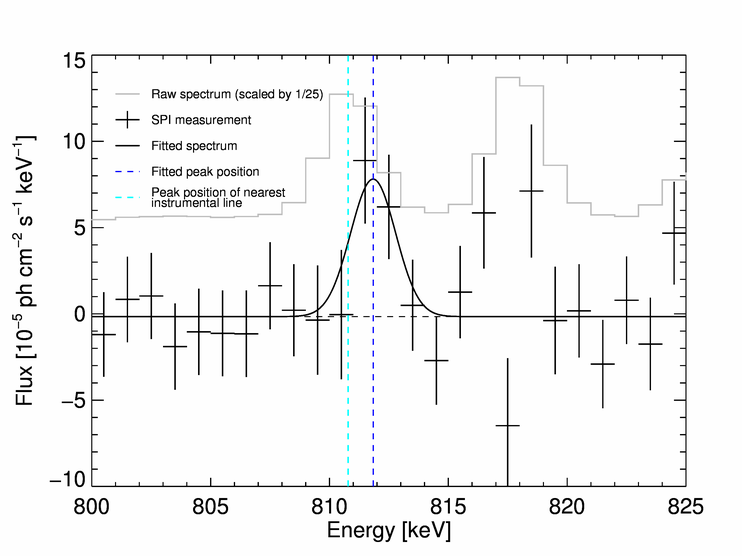}}
\caption{Spectra (black crosses) from the position of SN2014J between days 16.6 and 19.2 around 158 (left) and 812~keV (right), showing the two main $^{56}$Ni decay $\gamma$-ray lines. The solid black lines are Gaussians as fitted to the data points. The scaled raw spectrum is superimposed in gray, together with the peak positions of the fitted Gaussians compared to the peak positions of the nearest instrumental lines.
}
\vspace{-15pt}
\label{fig:line_raw_fits}
\end{figure}
\begin{wrapfigure}{r}{7cm}
\center
\vspace{-20pt}
\includegraphics[scale=0.3]{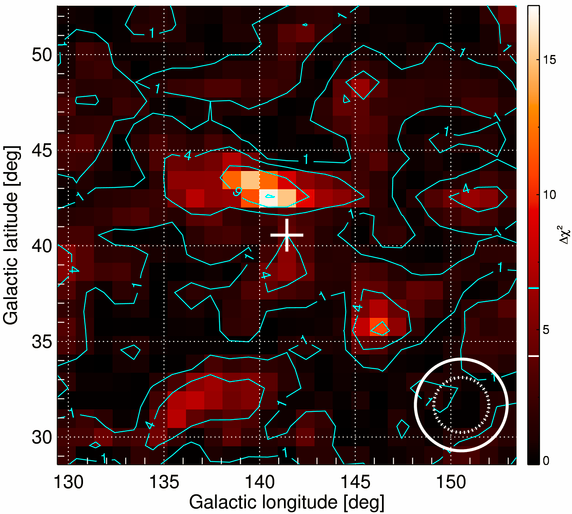}
\caption{Location of the $^{56}$Ni line signal on the sky. Within instrumental uncertainties (1$\sigma$ dashed, 2$\sigma$ solid circle), is the $\gamma$-ray emission from the 158 and 812~keV line mapped onto the position of SN2014J (cross).}
\vspace{-10pt}
\label{fig:signi_image}
\end{wrapfigure}

Figure~\ref{fig:line_raw_fits} shows the resulting spectra of a point source located at the position of SN2014J (at galactic coordinates $(l/b) = (141.427^{\circ}/40.558^{\circ})$) at days 16.6 to 19.2
after the explosion. The two major lines at 158 and 812~keV can well be described as Gaussians on top of a constant offset. Within the uncertainties, the two lines are identical. Also the imaging analysis locates the signal at the position of SN2014J, within spatial-resolution uncertainties, as shown in figure~\ref{fig:signi_image}.
Integrating the Gaussian line profiles yields the $\gamma$-ray line fluxes to be $(1.10\pm0.42)\cdot10^{-4}$~ph~cm$^{-2}$~s$^{-1}$ for the 158~keV line and $(1.90\pm0.66)\cdot10^{-4}$~ph~cm$^{-2}$~s$^{-1}$ for the 812~keV line. The fitted centroids for the two lines are $159.43\pm0.43$~keV and $811.84\pm0.42$~keV, respectively, consistent with the laboratory energies of $158.38$~keV and $811.85$~keV, thus corresponding to a bulk velocity below $\approx$2000~km~s$^{-1}$ (2$\sigma$ conf.). The measured intrinsic celestial line widths
suggest no significant Doppler-broadening. Although broader components could underlie the main signals, the velocity spreads are below 2000~km~s$^{-1}$ (2$\sigma$ conf.). The measured flux values can be converted to a $^{56}$Ni mass seen to decay at $\approx$18~days past explosion that corresponds to $(0.06\pm0.03)$~\Msol of $^{56}$Ni at the time of the explosion.\\
These values are rather small for typical scenarios in which $^{56}$Ni is buried deeply in the SN cloud where the all $^{56}$Ni $\gamma$-rays should have been absorbed at this time. Our measurement implies that a substantial fraction of $^{56}$Ni must have been produced close to the surface of the WD, at a depth not exceeding a few g~cm$^{-2}$ in column density.
\begin{figure}[Htbp]
\center
\vspace{-20pt}
\includegraphics[width=0.5\linewidth]{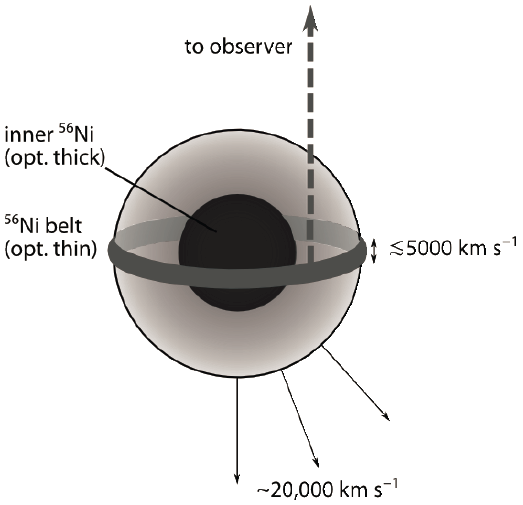}
\caption{Sketch of an ejecta configuration compatible with our observations. Helium accreted in a belt before the explosion produces the $^{56}$Ni belt at the surface of the ejecta. The $\gamma$-rays can escape from the belt material, while the $^{56}$Ni in the core (black) is still buried at high optical depths. The observer is believed to view the SN nearly face-on.}
\vspace{-10pt}
\label{fig:belt_config}
\end{figure}
Such an amount of $^{56}$Ni freely exposed, together with the constraints on kinematics, is unexpected in most explosion models.\\
Observations in other wavelengths also suggest SN2014J to be an unusual explosion, since the expected t$^2$-behaviour~\cite{Nugent2011_SN2011fe} in the rise of the optical lightcurve is not strictly represented by the data: The light curve hours after the explosion is rather well described by a broken powerlaw, showing a steeper rise and a possible shoulder which might also imply $^{56}$Ni near the surface~\cite{Goobar2014_SN2014J,Zheng2014_SN2014J}.\\
Classical SD scenarios have been excluded from upper limits of X-ray emission~\cite{Nielsen2014_SN2014J}. Sub-Chandrasekhar models with a He donating companion star or DD merger seem to work better in the case of SN2014J and already explain the $^{56}$Ni located near the surface~\cite{Ruiter2014_SNIa}. Most classical double-detonation models predict high velocity $\gamma$-ray lines to be observed if a shell of $^{56}$Ni was surrounding the SN ejecta~\cite{Fink2007_SNIa,Moll2013_SNIa,Fink2010_SNIa}. Moreover, this configuration should have an imprint on optical observables~\cite{Kromer2010_SNIa} which are not seen in SN2014J~\cite{Nielsen2014_SN2014J}. A plausible, though somewhat speculative configuration that explains the $\gamma$-ray data, is an equatorial He belt rather than a shell accreted from a He donor. If this situation would be observed essentially pole-on, our kinematic constraints can be met. Analogous to classical novae, where this idea has already been discussed~\cite{Kippenhahn1978_Novae,Piro2004_Novae}, unstable mass transfer on the Kelvin-Helmholtz time-scale from the He companion to the WD is possible if the He donor is more massive than the WD and fills its Roche lobe. In order to conserve a belt and not to spread the accreted He over the WD before the He flash explosion, either the accretion rate must be of the order 10$^{-5}$~\Msol~yr$^{-1}$~\cite{Law1983_SNIa} and/or the WD must be rapidly rotating so that the accretion is faster than it loses angular momentum. This high accretion rate can be achieved if the material is mostly He and not H. This would be the standard scenario for a supersoft X-ray source and is excluded for SN2014J~\cite{Greiner1996_binaries,Nielsen2014_SN2014J}. When this He belt becomes dense enough, explosive He burning may be ignited, and the configuration as shown in figure~\ref{fig:belt_config} can be established. Remarkably, this configuration is consistent with both, optical and $\gamma$-ray observation data. Moreover, our radiation transfer simulations~\cite{Diehl2014_SN2014J_Ni} do not produce easily distinguishable features in UVOIR wavelengths from the $^{56}$Ni-belt compared to standard models which makes SN2014J appear quite normal, at any observations angle.

\section{$^{56}$Co $\gamma$-ray measurements}
\begin{figure}[h!tbp]
\vspace{-20pt}
\subfigure{\includegraphics[width=0.49\textwidth]{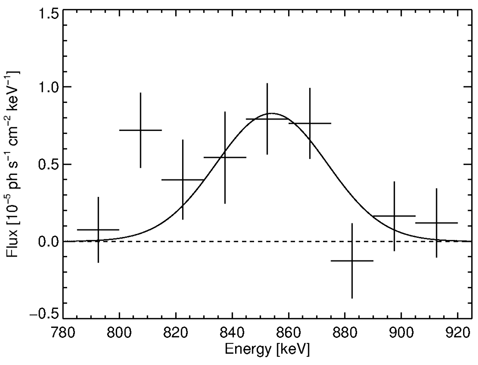}}\hfill
\subfigure{\includegraphics[width=0.49\textwidth]{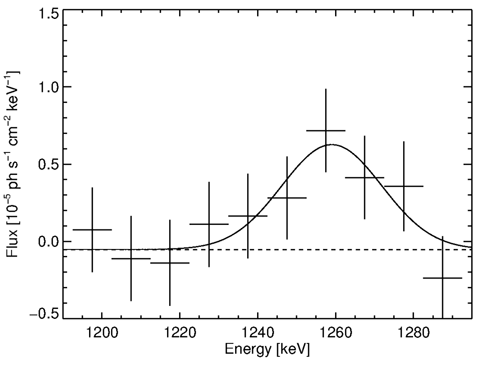}}
\caption{Spectra (black crosses) from the position of SN2014J at the expected $\gamma$-ray line intensity maximum between days $\approx$60 and $\approx$100 around 847 (left) and 1238~keV (right), showing the two main $^{56}$Co decay $\gamma$-ray lines. The solid black lines represent Gaussians as fitted to the dervied data points. Using the information from this broad binning, the lines appear broadened, equivalent to a velocity spread between 3000 and 8000~km~s$^{-1}$, and blue-shifted about 3000~km~s$^{-1}$. The measured line intensities are $(3.65\pm1.21)\cdot10^{-4}$~ph~cm$^{-2}$~s$^{-1}$ (847~keV) and $(2.27\pm0.69)\cdot10^{-4}$~ph~cm$^{-2}$~s$^{-1}$ (1238~keV), respectively, which corresponds to an initially synthesized $^{56}$Ni mass of $(0.50\pm0.12)$~\Msol, assuming the SN is transparent.}
\label{fig:line_fits_co}
\end{figure}
\begin{figure}[Htbp]
\subfigure{\includegraphics[trim=2cm 1cm 2cm 2cm,clip=true,width=0.33\textwidth]{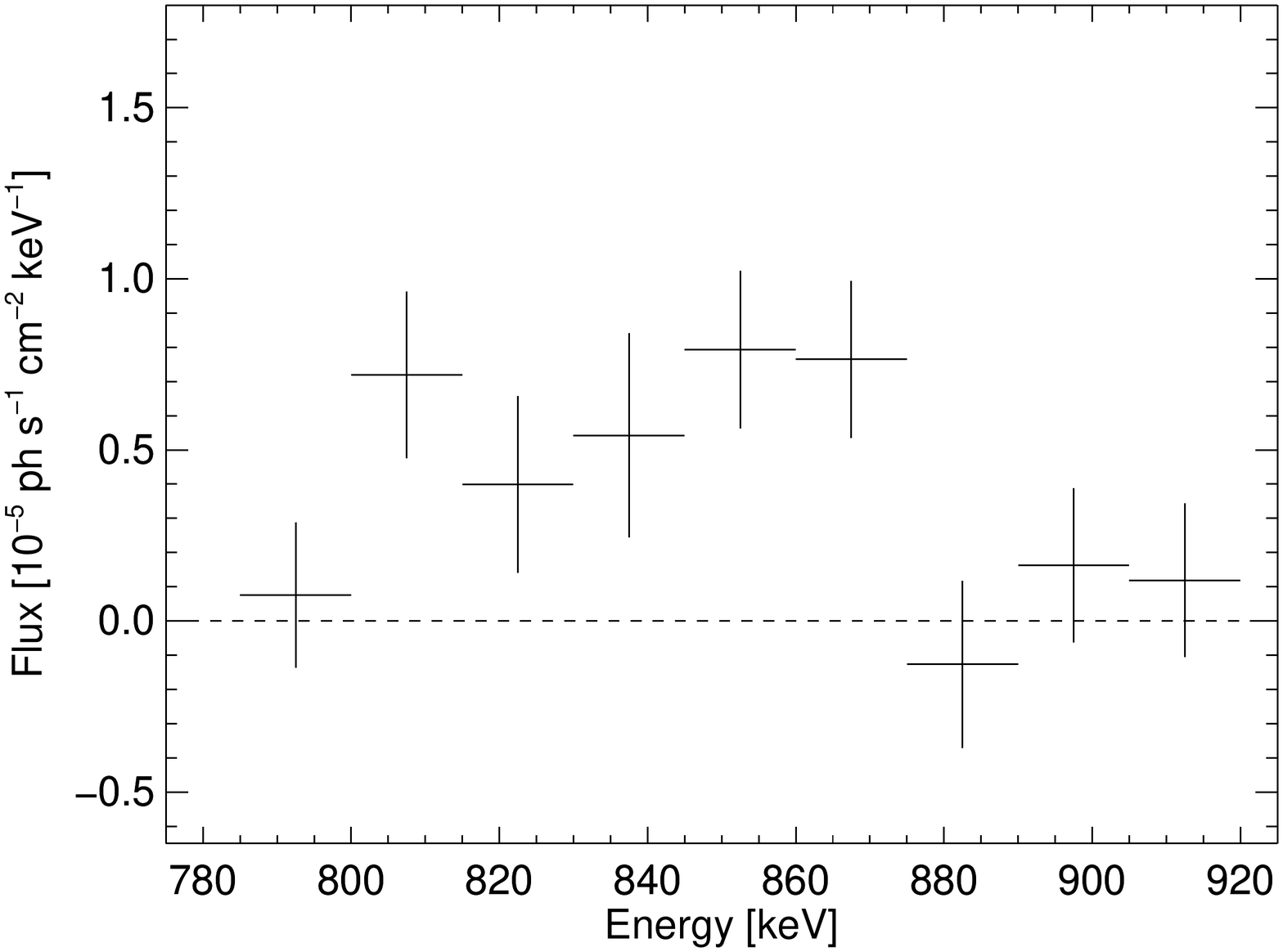}}
\subfigure{\includegraphics[trim=2cm 1cm 2cm 2cm,clip=true,width=0.33\textwidth]{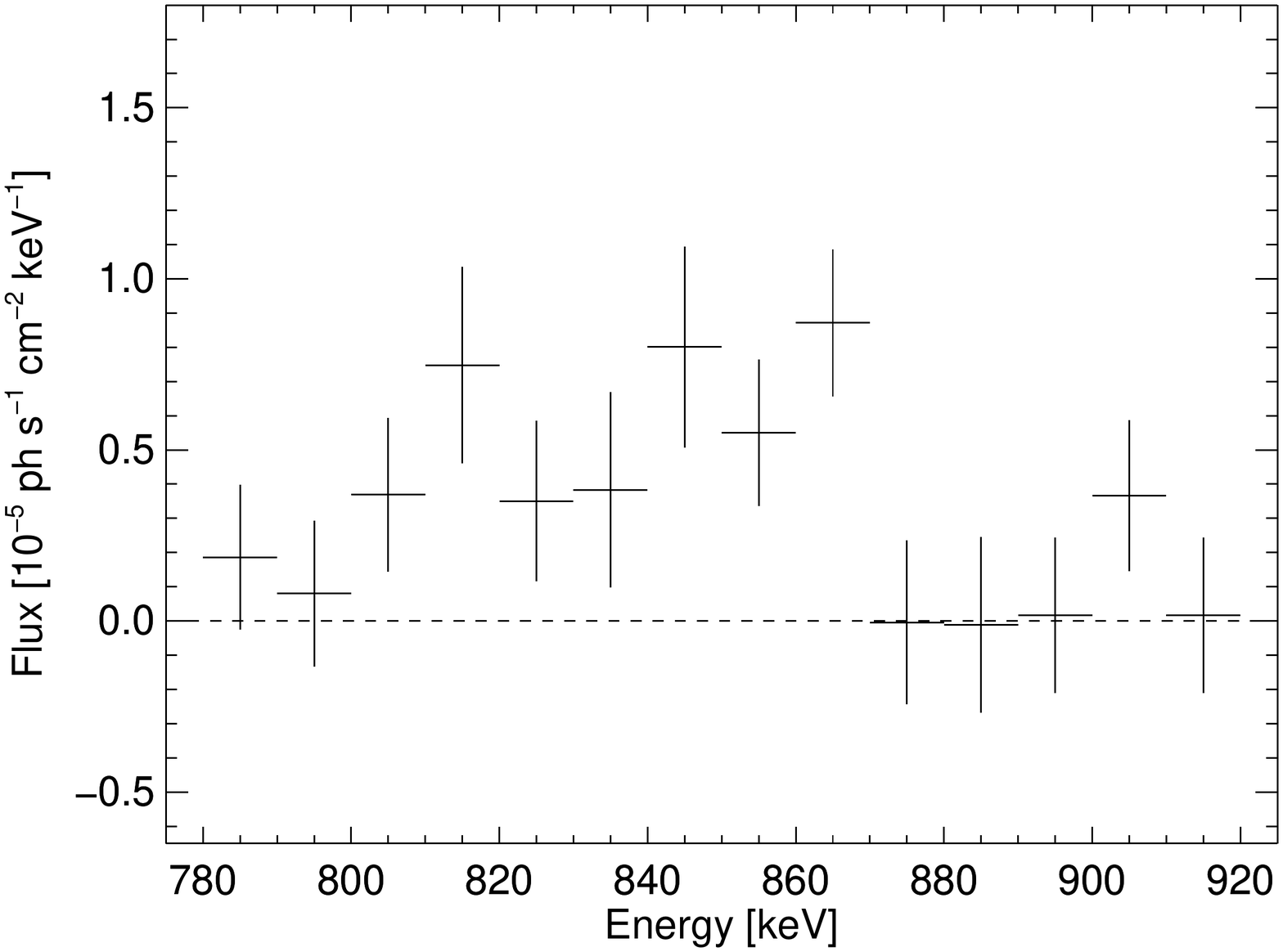}}
\subfigure{\includegraphics[trim=2cm 1cm 2cm 2cm,clip=true,width=0.33\textwidth]{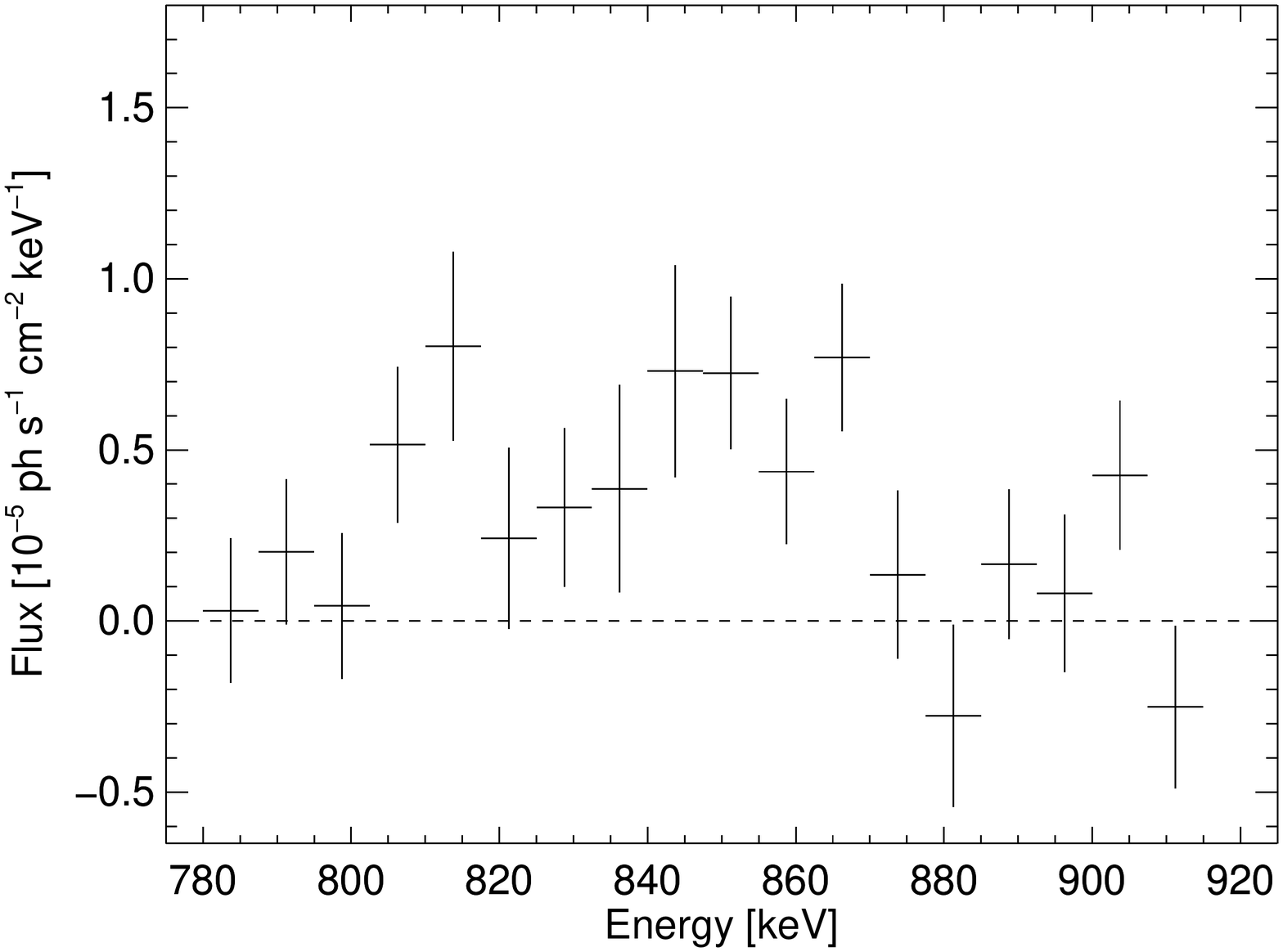}}\\
\subfigure{\includegraphics[trim=2cm 1cm 2cm 2cm,clip=true,width=0.33\textwidth]{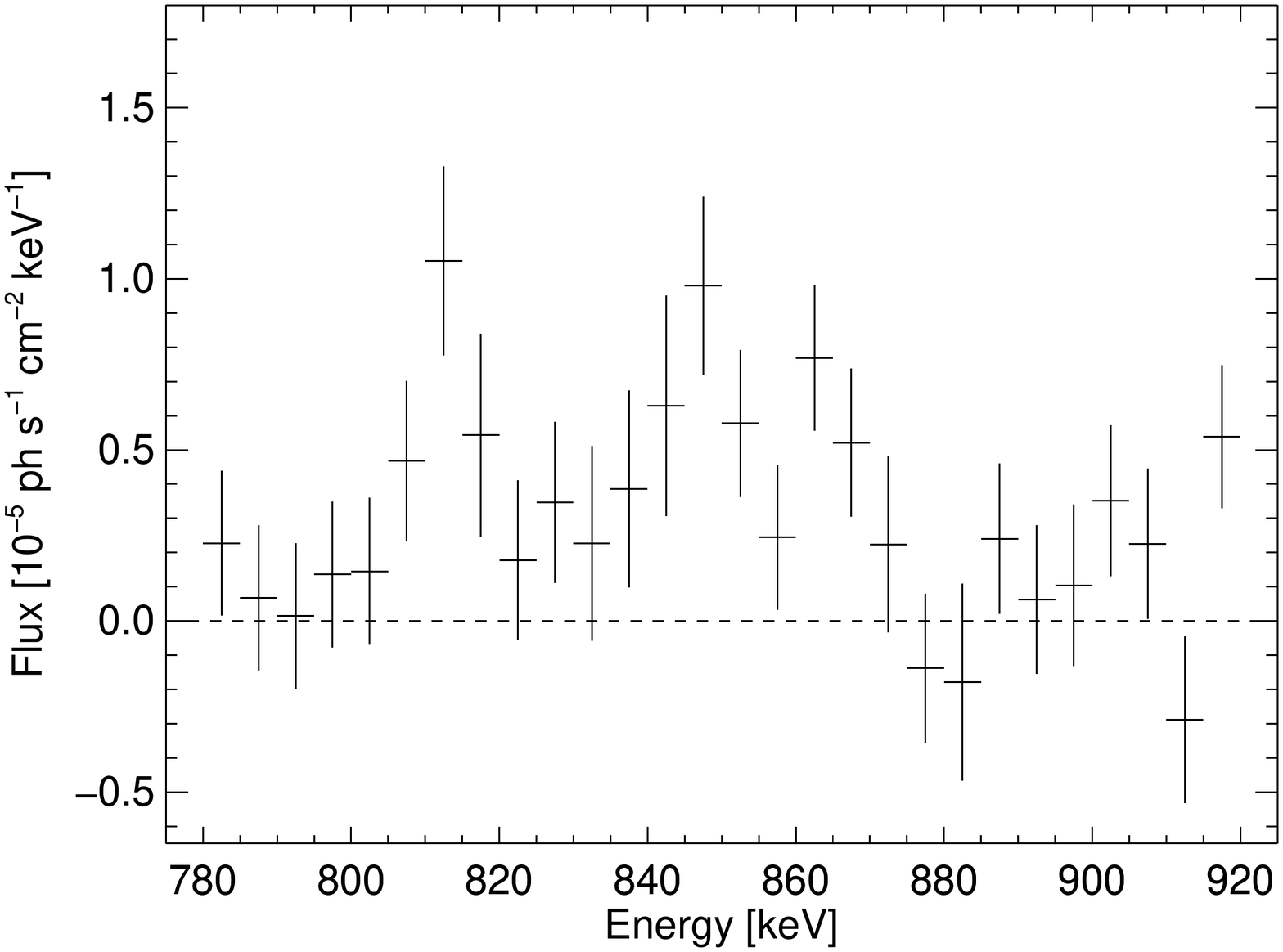}}
\subfigure{\includegraphics[trim=2cm 1cm 2cm 2cm,clip=true,width=0.33\textwidth]{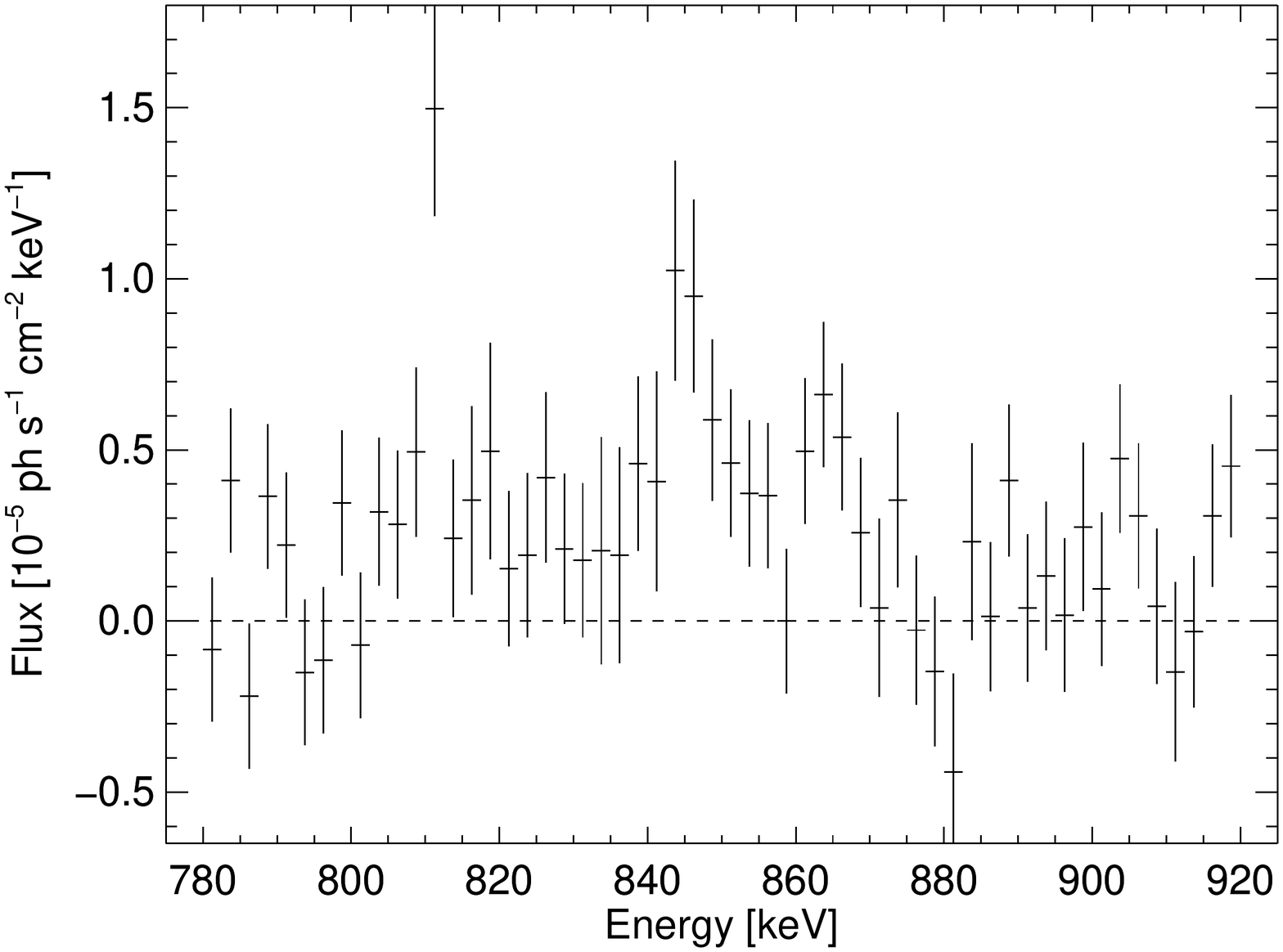}}
\subfigure{\includegraphics[trim=2cm 1cm 2cm 2cm,clip=true,width=0.33\textwidth]{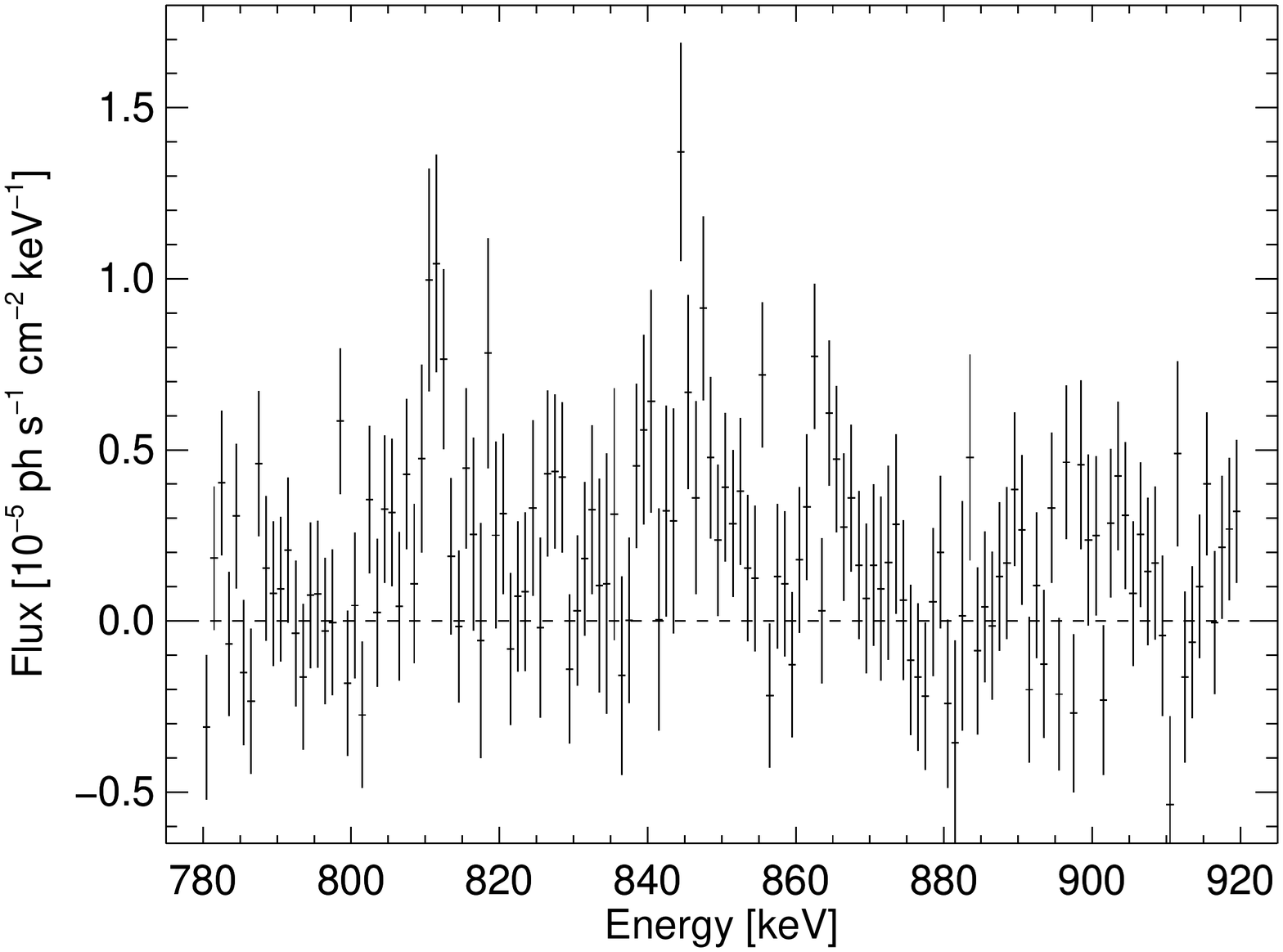}}
\caption{Series of the same spectrum as shown in figure~\protect\ref{fig:line_fits_co} (left panel) for different energy bin sizes. From top left to bottom right, the bin sizes are 15~keV, 10~keV, 7.5~keV, 5~keV, 2.5~keV, 1~keV. Accounting for the high resolution of SPI with its instrumental resolution of $\approx$2.3~keV at 850~keV, one single broad Gaussian is not representating the observed data any more. Apparently, at least three separate features can be identified and followed in time (see below).}
\vspace{-10pt}
\label{fig:spec_bin_evo}
\end{figure}
\begin{wrapfigure}{l}{8cm}
  \centering
  \includegraphics[width=1.0\linewidth]{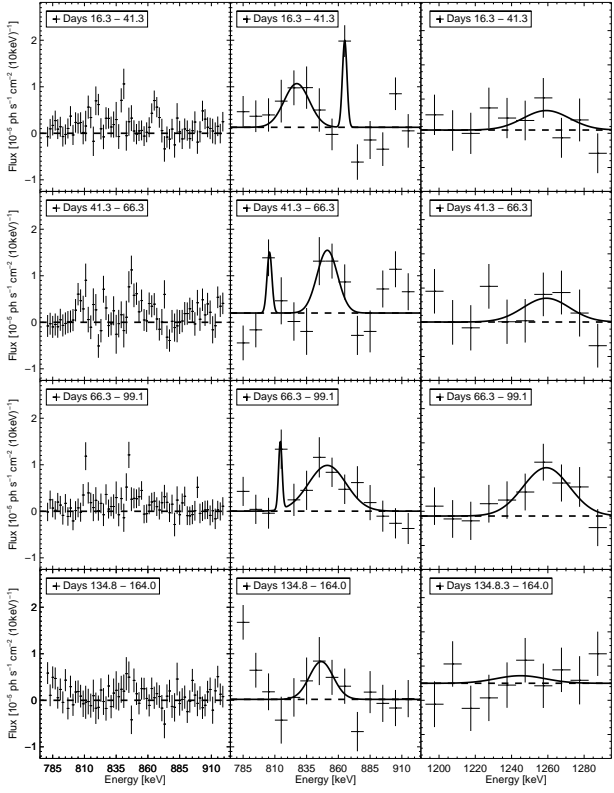}
  \caption{SN2014J signal intensity variations for the 847~keV line (center) and the 1238~keV line (right) as seen in the four chosen epochs, and analysed with 10~keV energy bins. The 1238~keV fluxes have been scaled by 1/0.68. Fitted line shape parameters are discussed in the text. For the 847~keV line, in addition, a high-spectral resolution analysis is shown at 2~keV energy in width (left).}
  \vspace{0pt}
  \label{fig:three_panel_spectra}
\end{wrapfigure}

Figure~\ref{fig:line_fits_co} shows the resulting spectra of a point source located at the position of SN2014J between days 66.3 and 99.1. 
Two $^{56}$Co $\gamma$-ray lines 
are clearly detected, and the flux ratio derived from the line intensities of $(3.65\pm1.21)\cdot10^{-4}$~ph~cm$^{-2}$~s$^{-1}$ from the 847~keV line and $(2.27\pm0.69)\cdot10^{-4}$~ph~cm$^{-2}$~s$^{-1}$ from the 1238~keV line is $0.62\pm0.28$ (lab value: 0.68). The mass of $^{56}$Ni derived from these flux values is $(0.50\pm0.12)$~\Msol, assuming the SN is fully transparent. This mass estimation, the constraints on the Doppler-broadening of about 40 to 50~keV (FWHM), corresponding to velocities of $(5500\pm2500)$~km~s$^{-1}$, and the constraints on the position of the lines to be blue-shifted about $(3000\pm1500)$~km~s$^{-1}$ are in good agreement with the majority of SN Ia model calculations.\\
However, our newly developed analysis method allows us to investigate the spectral shapes of the $^{56}$Co lines in even more detail. Figure~\ref{fig:spec_bin_evo} shows the same spectrum shown in figure~\ref{fig:line_fits_co} (left panel) at different energy binnings. Apparently, a single broad Gaussian does not capture the line shape properties any more, suggesting SN2014J to be an asymmetric explosion event.\\
In general, the $^{56}$Co $\gamma$-rays should emerge gradually from the inner part of the SN as its envelope gets more and more transparent over time while expanding~\cite{Isern1997_SNIa}. Thus, the spectral shapes at early times with respect to late times may be different, e.g. moving from a general blue-shift (near side) towards the laboratory energy. If ejecta would include major non-sphericities, velocity signatures could be more complex. Therefore, we separated our observations for both lines in four different epochs as shown in figure~\ref{fig:three_panel_spectra}: The first epoch, ranging from days 16.3 to 41.3 after the explosion, includes the optical light curve's maximum in which most $\gamma$-rays from the inner core should be absorbed. Epoch (2), between days 41.3 and 66.3 is the phase in which the $\gamma$-ray emission should gradually rise and end up at the maximum in the third epoch from days 66.3 to 99.1. After an observation gap between days 100 and 134, the fourth epoch (days 134.8 to 164.0) should capture the $\gamma$-ray emission in a rather transparent SN. The epochs are chosen to be sensitive to different SN Ia model scenarios, to help distinguish among them. Epoch (4) is almost independent of the early evolution and traces the global dynamics of the $^{56}$Co produced in the explosion. It should follow radioactive decay, in particular after the $\gamma$-ray maximum in a transparent SN.\\
Tracing the 847~keV feature at 10~keV binning in figure~\ref{fig:three_panel_spectra} (center panel), beginning in epoch (4), we identify a broadened line with a velocity spread of $(4570\pm1850)$~km~s$^{-1}$ at rest energy, fully consistent with the expectations. Towards earlier epochs (3) and (2), we find this line consistently, slightly blue-shifted and also broader.
Epoch (1), in contrast, shows a red-shifted ($\approx$7000~km~s$^{-1}$) broad line, together with a narrow blue-shifted feature. Which component corresponds to the bulk or peculiar $^{56}$Co emission is arbitrary, and can only be figured out by using smaller time bins, a higher spectral resolution and a more sophisticated explosion model: For the 847~keV line it was possible to divide the data set to 11 time bins. Using too short time intervals will drown any information about the spectral shapes in statistical fluctuations, and makes the detection of the different components difficult. The energy band from 780 to 920~keV was then analysed in 2~keV binning, resolving as narrow as the instrumental line width, for the 11 time bins independently. All earlier spectra show more than one spectral feature, and these are all less broad but combine to the broad features which are apparent when using coarser energy bins. The last three time bins, corresponding to epoch (4), show a consistent picture of a broad $^{56}$Co emission line on top of a rather flat spectrum in its vicinity. This reassures the strange behaviour at early times. Before day 99, $\approx$10~keV wide features emerge between 830 and 860~keV, varying in intensity and peak position with a consistent trend towards lower energies. At $\approx$810~keV, another line feature appears to rise and fall between days 60 and 90. Another feature at $\approx$865~keV which appears to be gone in the broad time and energy binning analysis seem to follow a radioactive $^{56}$Co decay. Such detailed discussion may over-interpret our data but we aim to point out a deviation from smoothness. We test the hypothesis of the early three epochs being statistical variations of the last, transparent, epoch by assuming the intrinsic line shape of epoch (4) and fit this template to the earlier epochs. The $\chi^2$-tests for epoch (2) and (3) reject this hypothesis at the $\approx2\sigma$ level. Especially in
epoch (1), the discrepancy is very large ($>4\sigma$), reassuring our finding of asymmetries in SN2014J. We also have used light curves of these three major spectral features to test a simple model incorporating individual $^{56}$Ni contents, kinematics and occultations. For the components, centered at 810, 845 and 865~keV, respectively, we find, consistently, early occulted and later revealed decaying $^{56}$Co line emission, and determine optical depths at day 1 of 2000, 600, and 100 for the three components (Maeda, priv. comm.).
The kinematics derived from this three-clump model are a blue-shifted $\approx$6400~km~s$^{-1}$ (at $\approx$865~keV), a red-shifted $\approx$13000~km~s$^{-1}$ (at $\approx$810~keV), and a main, spherical symmetric contribution (at $\approx$845~keV) to the total $^{56}$Ni mass. The relative parts of each feature are 15\%, 30\% and 55\%, respectively. We do not propose this to be the real morphology of SN2014J, but rather mention that this analysis supports our interpretation of an aspherical $^{56}$Ni distribution due to the differences in the respective occultation of the major features by outer SN material. Clumps or co-moving volume elements carrying $^{56}$Co may lie along less-occulted lines of sight at specific times. For this reason, as the SN expands, different volume elements may thus contribute at different times, as long as occulation of the $\gamma$-rays is significant.\\
\begin{figure}[H!tbp]
  \centering
  \includegraphics[width=0.7\linewidth]{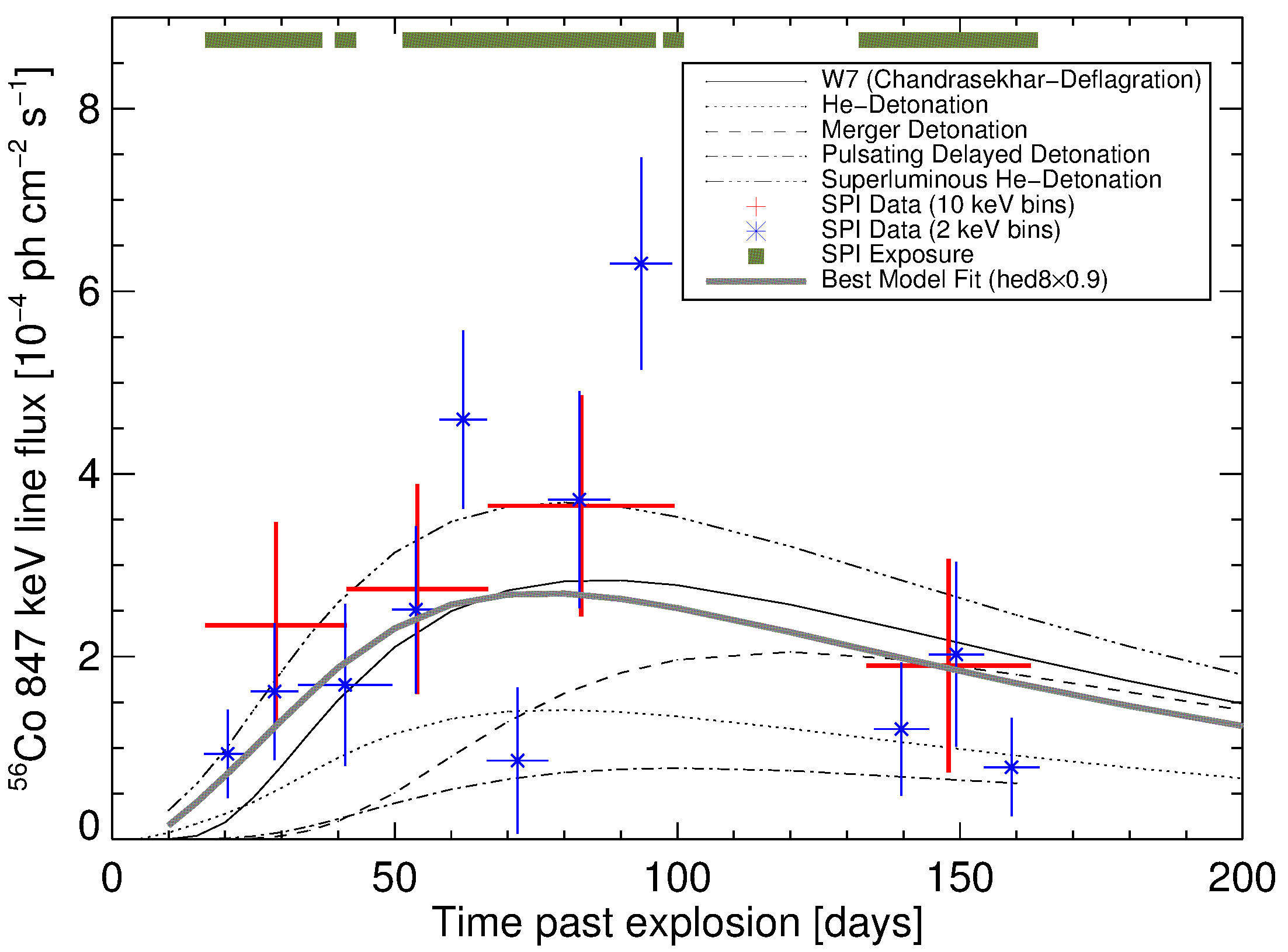}
  \caption{SN2014J signal intensity variations for the 847~keV line in two different time resolutions. The 4-epoch results are consistent with 11-epoch analysis, both showing an initial rise and late decline of $^{56}$Co decay line intensity, with a maximum at 60 to 100 days after the explosion. Shown are also several candidate source models from~\protect\cite{The2014_SN2014J}, fitted in intensity and thus determining $^{56}$Ni masses for each such model. The best-fitting model is shown as a continuous thick line.}
  \label{fig:co_light_curve}
\end{figure}
In a last step 
we compare the light curve of our brightest $^{56}$Co feature at 847~keV (see figure~\ref{fig:co_light_curve}) to models, prepared by Lih-Sin The and Adam Burrows~\cite{The2014_SN2014J} specifically for SN2014J, and use three different approaches to determine the initially synthesized $^{56}$Ni mass. Details about the procedure can be found in~\cite{Diehl2014_SN2014J_Co}.
The $^{56}$Ni is then determined by a template fit of the models by~\cite{The2014_SN2014J}, with a global scaling parameter, corresponding to the $^{56}$Ni mass.
The most plausible models give a $^{56}$Ni mass in the range 0.46-0.59 $(\pm0.06)$~\Msol, and all imply a substantial fraction of $^{56}$Ni near the outskirts of the SN. We note that distinguishing between different models from their fit-quality is difficult as the $\chi^2$ is not varying very much for most of the models.
The detailed summary of the fitting results can be found in~\cite{Diehl2014_SN2014J_Co}.
Taking the average of the three best-fitting models in all three approaches yields a $^{56}$Ni mass of $(0.49\pm0.09)$~\Msol. In general, the values range between 0.4 and 0.8~\Msol. The preference of specific models suggests He on the outside of the SN. These (1D) models assume a single broad line to create their time profiles, which of course do not trace the details in the flux measurements, even for the best-fitted models. If the spectral constraints are relaxed, the light curve fits improve, and hence are in good agreement with our proposed irregular appearence of the $^{56}$Co lines in SN2014J. We note that in spite of clear detections of the $^{56}$Co $\gamma$-ray lines, discriminating among models is still inadequate, and most models are in general agreement with our measurements if $^{56}$Ni is a free parameter. A general model sophistication including different ejecta morphologies, radiation transport and a 3D approach would be in order, to tighten the constraints on possible progenitor systems and explosion scenarios.

\section{Conclusions}
We analysed the full set of INTEGRAL/SPI observations available for SN2014J and concentrated on the energy ranges around the brightest $\gamma$-ray line emission expected from the $^{56}$Ni decay chain at 158~keV, 812~keV, 847~keV, and 1238~keV. The large instrumental BG of SPI is modelled by a detailed spectral analysis which exploits the instrumental spectral response, and accounts for continuum and many instrumental $\gamma$-ray lines arising from cosmic-ray interactions with the instrument and spacecraft, in a broader energy range around the lines of interest. In this new BG modeling approach, we first identify long-term spectral features, then separately fit their appearances in each individual Ge detector with its response characteristics, and finally determine the amplitude variations of the spectral background template on the shorter time scales of our individual pointings. This method overcomes limitations of statistical precision of BG measurements in the fine energy bins where we wish to analyse the source, and uses BG data as measured by the SPI detectors themselves, and the nuclear reaction physics extracted from long-term monitoring of BG and detector response (Siegert et al., in prep.).\\
Our measurements of $^{56}$Ni decay in SN2014J find the two strongest $\gamma$-ray lines at 158 and 812~keV at day 20 after the explosion. This is surprising not only because of the detection itself but due to the kinematics derived. Since most of the $^{56}$Ni $\gamma$-rays should be absorbed in the SN cloud at that time, a major amount of $^{56}$Ni must have been produced outside the SN. The constraints on the measured low Doppler-shifts and -velocities indicate an asymmetry in SN2014J. One plausible explanation may be a belt of He accreted from a companion star which is exploding, producing $^{56}$Ni on the surface of the WD, and triggering the SN.\\
Moreover, we find emission of decaying $^{56}$Co during the whole observation period of SN2014J at 847 and 1238~keV, respectively. We analyse separately different epochs of our observations, and derive an intensity time profile in the energy regime around the two lines. The brightness evolution is consistent for both lines, and is overall consistent with expectations from different plausible explosion models. We find irregularities in the appearance of the line emission during the rise of the measured SN Ia $\gamma$-ray light-curve which can not be described as a smooth emergence of the Doppler broadened emission lines as they are found in the late, $\gamma$-ray transparent
phase. The derived spectra indicate transient emission features of different widths which is hard to assign clearly to $^{56}$Ni or $^{56}$Co. The three-dimensional structure of the SN does not follow the regular shape as predicted by one-dimensional models.\\
We note that our data cannot discriminate among 1D models, although our measurements consistently favour models in which He is on the outskirts. In both measurements we find that the explosion SN2014J was not spherically symmetric, dervied from the observed line shapes. These results demonstrate the power of $\gamma$-ray line observations from SNe Ia, and seed high expectations should a closer event occur during the ongoing INTEGRAL mission.

\bibliographystyle{bibgen}

\begin{thebibliography}{10}

\bibitem{Goobar2011_SNIa}
A.~{Goobar} \& B.~{Leibundgut}. Nov. 2011, Annual Review of Nuclear and
  Particle Science, 61, 251

\bibitem{Hillebrandt2000_SNIa}
W.~{Hillebrandt} \& J.~C. {Niemeyer}. 2000, \araa, 38, 191

\bibitem{Li2011_SNIa}
W.~{Li}, J.~S. {Bloom}, et~al. Dec. 2011, \nat, 480, 348

\bibitem{Hillebrandt2013_SNIa}
W.~{Hillebrandt}, M.~{Kromer}, et~al. Apr. 2013, Frontiers of Physics, 8, 116

\bibitem{Guillochon2010_SNIa}
J.~{Guillochon}, M.~{Dan}, et~al. Jan. 2010, \apjl, 709, L64

\bibitem{Rosswog2009_SNIa}
S.~{Rosswog}, D.~{Kasen}, et~al. Nov. 2009, \apjl, 705, L128

\bibitem{Fink2010_SNIa}
M.~{Fink}, F.~K. {R{\"o}pke}, et~al. May 2010, \aap, 514, A53

\bibitem{Pakmor2012_SNIa}
R.~{Pakmor}, M.~{Kromer}, et~al. Mar. 2012, \apjl, 747, L10

\bibitem{Nomoto1997_SNIa}
K.~{Nomoto}, K.~{Iwamoto}, et~al. Feb. 1997, Nuclear Physics A, 621, 467

\bibitem{Mazzali2007_SNIa}
P.~A. {Mazzali}, F.~K. {R{\"o}pke}, et~al. Feb. 2007, Science, 315, 825

\bibitem{Arnett1982_SNIa}
W.~D. {Arnett}. Feb. 1982, \apj, 253, 785

\bibitem{Stritzinger2006_SNIa}
M.~{Stritzinger}, B.~{Leibundgut}, et~al. Apr. 2006, \aap, 450, 241

\bibitem{Isern2008_SNIa}
J.~{Isern}, E.~{Bravo}, et~al. Oct. 2008, \nar, 52, 377

\bibitem{Roepke2012_SNIa}
F.~K. {R{\"o}pke}, M.~{Kromer}, et~al. May 2012, \apjl, 750, L19

\bibitem{Bours2013_SNIa}
M.~C.~P. {Bours}, S.~{Toonen}, et~al. Apr. 2013, \aap, 552, A24

\bibitem{Ruiz2004_Tycho}
P.~{Ruiz-Lapuente}, F.~{Comeron}, et~al. Oct. 2004, \nat, 431, 1069

\bibitem{Kerzendorf2009_Tycho}
W.~E. {Kerzendorf}, B.~P. {Schmidt}, et~al. Aug. 2009, \apj, 701, 1665

\bibitem{Schaefer2012_SNR0509}
B.~E. {Schaefer} \& A.~{Pagnotta}. Jan. 2012, \nat, 481, 164

\bibitem{Bloom2012_SN2011fe}
J.~S. {Bloom}, D.~{Kasen}, et~al. Jan. 2012, \apjl, 744, L17

\bibitem{Shappee2013_SN2011fe}
B.~J. {Shappee}, K.~Z. {Stanek}, et~al. Jan. 2013, \apjl, 762, L5

\bibitem{Goobar2014_SN2014J}
A.~{Goobar}, J.~{Johansson}, et~al. Mar. 2014, \apjl, 784, L12

\bibitem{Nomoto1984_SNIa}
K.~{Nomoto}, F.~K. {Thielemann}, et~al. Aug. 1984, \apjl, 283, L25

\bibitem{Khokhlov2001_SNIa}
A.~M. {Khokhlov}. 2001, Supernovae and Gamma-Ray Bursts: the Greatest
  Explosions since the Big Bang, eds. M.~{Livio}, N.~{Panagia}, \& K.~{Sahu},
  239--249

\bibitem{Woosley1994_SNIa}
S.~E. {Woosley} \& T.~A. {Weaver}. Mar. 1994, \apj, 423, 371

\bibitem{Livne1995_SNIa}
E.~{Livne} \& D.~{Arnett}. Oct. 1995, \apj, 452, 62

\bibitem{Isern2011_SNIa}
J.~{Isern}, M.~{Hernanz}, et~al. 2011, Lecture Notes in Physics, Berlin
  Springer Verlag, eds. R.~{Diehl}, D.~H. {Hartmann}, \& N.~{Prantzos}, vol.
  812 of Lecture Notes in Physics, Berlin Springer Verlag, 233--308

\bibitem{Milne2004_SNIa}
P.~A. {Milne}, A.~L. {Hungerford}, et~al. Oct. 2004, \apj, 613, 1101

\bibitem{Dessart2014_SNIa}
L.~{Dessart}, S.~{Blondin}, et~al. Jun. 2014, \mnras, 441, 532

\bibitem{The2014_SN2014J}
L.-S. {The} \& A.~{Burrows}. May 2014, \apj, 786, 141

\bibitem{Vedrenne2003_SPI}
G.~{Vedrenne}, J.-P. {Roques}, et~al. Nov. 2003, \aap, 411, L63

\bibitem{Winkler2003_INTEGRAL}
C.~{Winkler}, T.~J.-L. {Courvoisier}, et~al. Nov. 2003, \aap, 411, L1

\bibitem{Cao2014_SN2024J}
Y.~{Cao}, M.~M. {Kasliwal}, et~al. Jan. 2014, The Astronomer's Telegram, 5786,
  1

\bibitem{Fossey2014_SN2014J}
J.~{Fossey}, B.~{Cooke}, et~al. Jan. 2014, Central Bureau Electronic Telegrams,
  3792, 1

\bibitem{Foley2014_SN2014J}
R.~J. {Foley}, O.~D. {Fox}, et~al. Oct. 2014, \mnras, 443, 2887

\bibitem{Kuulkers2014_SN2014J}
E.~{Kuulkers}. Jan. 2014, The Astronomer's Telegram, 5835, 1

\bibitem{Zheng2014_SN2014J}
W.~{Zheng}, I.~{Shivvers}, et~al. Mar. 2014, \apjl, 783, L24

\bibitem{Diehl2014_SN2014J_Ni}
R.~{Diehl}, T.~{Siegert}, et~al. Sep. 2014, Science, 345, 1162

\bibitem{Diehl2014_SN2014J_Co}
R.~{Diehl}, T.~{Siegert}, et~al. Sep. 2014, ArXiv e-prints

\bibitem{Weidenspointner2003_SPI}
G.~{Weidenspointner}, J.~{Kiener}, et~al. Nov. 2003, \aap, 411, L113

\bibitem{Kretschmer2011_PhD}
{{Kretschmer}, K.~A.} 2011, Dissertation, {Technische Universit\"at M\"unchen},
  {M\"unchen}

\bibitem{Nugent2011_SN2011fe}
P.~E. {Nugent}, M.~{Sullivan}, et~al. Dec. 2011, \nat, 480, 344

\bibitem{Nielsen2014_SN2014J}
M.~T.~B. {Nielsen}, M.~{Gilfanov}, et~al. Aug. 2014, \mnras, 442, 3400

\bibitem{Ruiter2014_SNIa}
A.~J. {Ruiter}, K.~{Belczynski}, et~al. May 2014, \mnras, 440, L101

\bibitem{Fink2007_SNIa}
M.~{Fink}, W.~{Hillebrandt}, et~al. Dec. 2007, \aap, 476, 1133

\bibitem{Moll2013_SNIa}
R.~{Moll} \& S.~E. {Woosley}. Jan. 2013, American Astronomical Society Meeting
  Abstracts \#221, vol. 221 of American Astronomical Society Meeting Abstracts,
  \#253.22

\bibitem{Kromer2010_SNIa}
M.~{Kromer}, S.~A. {Sim}, et~al. Aug. 2010, \apj, 719, 1067

\bibitem{Kippenhahn1978_Novae}
R.~{Kippenhahn} \& H.-C. {Thomas}. Feb. 1978, \aap, 63, 265

\bibitem{Piro2004_Novae}
A.~L. {Piro} \& L.~{Bildsten}. Aug. 2004, \apj, 610, 977

\bibitem{Law1983_SNIa}
W.~Y. {Law} \& H.~{Ritter}. Jun. 1983, \aap, 123, 33

\bibitem{Greiner1996_binaries}
J.~{Greiner}, ed. 1996, {Supersoft X-Ray Sources}, vol. 472 of Lecture Notes in
  Physics, Berlin Springer Verlag

\bibitem{Isern1997_SNIa}
J.~{Isern}, J.~{G{\'o}mez-Gomar}, et~al. 1997, The Transparent Universe, eds.
  C.~{Winkler}, T.~J.-L. {Courvoisier}, \& P.~{Durouchoux}, vol. 382 of ESA
  Special Publication, 89

\end{thebibliography}

\end{document}